\documentclass[prb,aps,twocolumn,showpacs,amsmath,amssymb]{revtex4}
\usepackage{epsf}

\begin{document}
\draft
\newcommand{\ve}[1]{\boldsymbol{#1}}

\title{Electrostatic interface tuning in correlated superconducting heterostructures}
\author{Natalia Pavlenko and Thilo Kopp}
\address{Institute of Physics, University of Augsburg, 86135 Augsburg, Germany}

\begin{abstract}
An electrostatic field, which is applied to a gated
high-temperature superconducting (HTSC) film, is believed to affect
the film  similar to charge doping. Analyzing the pairing in terms of a
$t$--$J$ model, we show
that a coupling to electric dipoles and phonons at the
interface of film and dielectric gate localizes the injected charge and
leads to a superconductor-insulator transition.
This results in a dramatic modification of the doping dependent phase
diagram close to and above the optimal doping which is expected to
shed light on recent electric field-effect experiments with
HTSC cuprates.
\end{abstract}


\maketitle \section{Introduction}
Interface physics of strongly correlated
oxides is a rapidly developing branch of materials science. In heterostructures
of ultrathin correlated oxide films, charge and spin states are reconstructed
at the interfaces and hence affect the electronic properties of the entire
system \cite{ahn,ohtomo}. This interface-controlled behavior provides new
opportunities for oxide-film electronics, where a powerful tool for tuning the
heterostructure properties is the electric field \cite{ahn,zhao}. In
high-temperature superconducting oxides, electric fields can be used to switch
between superconducting and insulating states by electrostatically tuning the
free carrier density \cite{mannhart,ahn2}. In contrast to chemical doping,
where the modifications in the doping level are inevitably related to changes
in chemical bonding and microstructure, the field-effect experiments are
expected to only modify charge, keeping the microstructure fixed.

In superconducting field-effect transistors (SuFET's), an electric field is
applied to a dielectric gate and sweeps charge carriers into a HTSC-film
in the drain-source channel. The gate polarization attracts the injected
carriers at the gate/film-interface. It thereby creates an accumulation region
with a shifted local $T_c$ which, for a nm-thick film, should result in a
tuning of the global $T_c$. With the assumption of a fixed interface microstructure,
the field-induced $T_c$ shift should be fully determined by the electric field
doping. However, whereas in underdoped HTSC cuprate films the observed $T_c$
shift is about $5$--$18$~K, in the overdoped films $T_c$ is not substantially
changed by the field \cite{mannhart,ahn2}.
This striking doping dependence is usually
explained in terms of a Thomas-Fermi screening length
$\lambda_{TF}$ which suggests a stronger exponential decay of
accumulated charge inside the overdoped films with higher total
carrier density. However, in anisotropic systems, the quantum
mechanically determined spread of the charge from the interface is greater and
decays slower, in distinction to the classical Thomas-Fermi
approach \cite{ando}. As the cuprate films contain CuO$_2$-planes with an interplanar
distance $\sim 1$~nm, only a particular amount of injected charge
is accumulated in the first plane at the interface, and the rest
is redistributed in the $2$--$4$ nearest planes \cite{wehrli}. In these
multilayers, the charge confinement to the first plane increases for higher total charge
densities in the film \cite{wehrli}. Consequently, in the underdoped cuprate film,
despite larger $\lambda_{TF}$, the low carrier density results in
$\sim 80\%$ of confined charge (mechanisms to
achieve stronger confinement have been studied in
Ref.~\onlinecite{pavlenko2}). In the overdoped film, about $100
\%$ of the injected charge is confined \cite{wehrli}. As in both
cases the calculated field-accumulated charge densities are
roughly equal, one should expect similar shifts of $T_c$ {\it both
in the under- and overdoped films}. This similarity implies that
without consideration of the microscopic processes at the gate/film interface, the field doping
alone cannot satisfactorily explain the
doping dependences of the $T_c$ shift.

In a typical SuFET, the strong modulation of injected
carriers requires a gate polarization in the range
$10$--$30\mu$~C/cm$^2$. To achieve this polarization, one uses gates with
dielectric constant $\epsilon$ in the range $20$--$100$,
fabricated from complex perovskite transition metal oxides like
SrTiO$_3$ (STO). Here, extensive studies
\cite {migoni} show that a fundamental property is the hybridization of oxygen (O) $p$- and
transition metal (Ti) $d$-orbitals caused by the Ti--O-displacements. The Coulomb interaction
$V_{pd}$ of a small amount of injected holes with
the hybridized $p$-$d$ electrons in the insulating gate remains
almost unscreened, $V_{pd}\sim 1$--$2$~eV---the spatial distance
between the nearest interface unit cells of gate and film is about
$2.5$--$5$~\AA. Moreover, the holes strongly interact with  optical
TiO-phonons which polarize the interface. These excitation processes
at the interface can possibly lead to charge localization.
A further reason for interface charge trapping can be an increased disorder
at the cuprate/STO-interface which can lead to Anderson localization.
In fact, interface localization is experimentally supported by a
hysteresis of the normal state resistance and by the voltage
dependence of the hole mobility in HTSC/STO-heterostructures \cite{mannhart,eckstein}.
There are also clear experimental indications that in organic field effect transistors, where
a similar mechanism of the electric field effect applies, the carrier mobility decreases
with increasing $\epsilon$. \cite{stassen}

As a step towards an understanding of the cooperative interface phenomena in correlated oxides,
we present a theoretical study of field-doped heterostructures,
whereby we focus on two key aspects: (i) the interaction of the
injected charge carriers with $p$-$d$-electrons
in the dielectric gate, and (ii) the coupling of carriers to dynamical lattice
distortions. We find that these processes at the gate/film
interface result in dramatic modifications of the superconducting state
with increasing field-induced carrier density.

\section{Microscopic scheme of the electric field effect}
\begin{figure}[t]
\epsfxsize=6.5cm \centerline{\epsffile{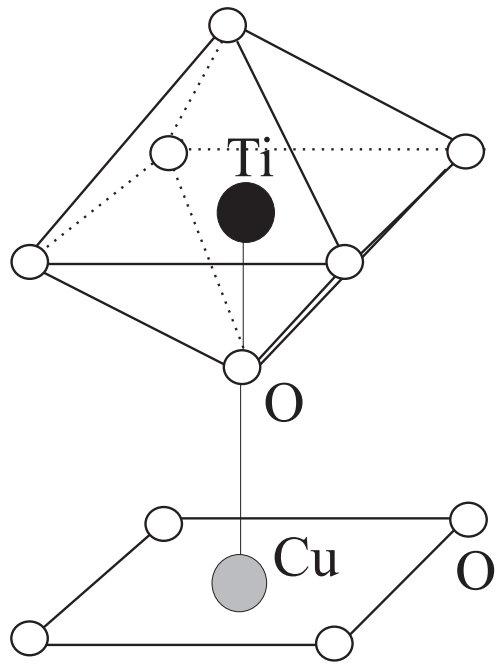}} \caption{Schematic
presentation of a possible interface bonding in HTSC/STO heterostructure. The
TiO group of TiO$_6$-octahedron which is directed perpendicular to the
interface, is bonded to the Cu$^{2+}$ ion of the interface CuO$_2$-plane of the
cuprate superconducting film.} \label{fig1}
\end{figure}

To analyze the field dependence of the $p$-$d$ hybridization at the gate
interface, we introduce a model on a 2D-square lattice containing $N_{\perp}$
sites, where each site corresponds to a perovskite unit cell. In each
O$_6$-octahedron of the $i$-th cell we consider a single TiO ionic group with a
dynamical covalent bond along the field direction perpendicular to the
gate/film-interface (shown in Fig.~\ref{fig1}). In this TiO group, we include
only two states ($p$ and $d$) expressed by the operators $d_i^{\dag}$($d_i$)
and $p_i^{\dag}$($p_i$) which obey the one-electron constraint
$d_i^{\dag}d_i+p_i^{\dag}p_i=1$. In a SuFET, the gate electric field
$\varepsilon_g$ affects the $p$-$d$-electron transfer:
\begin{equation}\label{dg1}
H_{pd}=\frac{1}{2} \Delta^0_{pd} \sum_{i} (d_i^{\dag}d_i-p_i^{\dag}p_i)+
  E_{pd}\sum_i (p_i^{\dag}d_i+d_i^{\dag}p_i).
\end{equation}
The energy gap $\Delta^0_{pd}$ is about $3$~eV in perovskite dielectric states,
but decreases down to $1.5$~eV in ferroelectric films due to modifications of
surface electronic states \cite{krcmar}. The electrostatic interaction
$E_{pd}=\varepsilon_g d_{pd}$ of the gate field with the dipole moment,
$d_{pd}=e r_{pd}$, of the electron transfer increases the $p$-$d$ level
splitting and thus tends to localize the TiO-electron. Here $e$ is the electron
charge and $r_{pd}\approx 2$~\AA ~is the Ti-O distance. The energy gap
$\Delta^0_{pd}$ in (\ref{dg1}) refers to $\varepsilon_g=0$. In a
nonzero electric field, which induces a polarization $P_{\varepsilon}\sim
\varepsilon_g$, the field-dependent TiO-distortions $u_i\sim P_{\varepsilon}$ imply
$\Delta_{pd}^{\varepsilon}=\Delta^0_{pd}+\chi_{\varepsilon} u_i$. The field effect can be
taken into account via the expansion of $\Delta_{pd}^{\varepsilon}$ in
powers of $P_{\varepsilon}$:
$\Delta_{pd}^{\varepsilon}=\Delta_{pd}^{\varepsilon}(P_S^0+\frac{\partial P_S^0}{\partial
\varepsilon_g} \varepsilon_g)$, which determines the "susceptibility" 
$\chi_{\varepsilon} \sim  \frac{\partial \Delta_{pd}^0}{\partial P_S} \frac{\partial
P_S^0}{\partial \varepsilon_g}$. In thin STO films, the high concentration
of oxygen vacancies leads to local polar regions and to a quasistatic
polarization $P_S^0 \ne 0$.~\cite{xi3} The additional, maximal electric
field-induced polarization is comparable to $P_S^0$ and the influence of
$\varepsilon_g$ on $P_S$ is relatively weak compared to the effect produced by
the polar regions. Therefore, for the considered range of $\varepsilon_g <
10^6$~V/cm, we have $\chi_{\varepsilon} \varepsilon_g \ll P_S^0$ and consequently we take
$\Delta_{ps}^{\varepsilon}=\Delta^0_{pd}$ in (\ref{dg1}).

Furthermore, we assume that the barrier for a $p$-$d$-transfer (described by
the second term in (\ref{dg1})) is modified by the interface coupling to the
carriers in the superconducting film
\begin{equation}\label{dg_s}
H_{{\rm exc}}= V_{pd}\sum_{i\sigma} (1-n_{i\sigma}) (p_i^{\dag}d_i+d_i^{\dag}p_i).
\end{equation}
We consider in (\ref{dg_s}) the coupling with holes,
where $n_{i\sigma}$ is the electron number operator with spin $\sigma$,
and $n_i=\sum_{\sigma}n_{i\sigma}$.
The mechanism (\ref{dg_s}) can lead to a local effective attraction
between the injected carriers, which has 
recently been analyzed for weak-coupling $s$-wave superconductors \cite{kopp}.

The role of the TiO-distortions is more subtle.
As shown in Ref.~\onlinecite{pavlenko}, the coupling of the injected charge to a static displacement
only shifts the chemical potential in the film. Consequently, we neglect the coupling to
static distortions and we focus instead on dynamical
TiO-displacements. We are interested in the coupling to the
low-energy soft TO$_1$-mode, the polar component of which is hardened up to
$50$--$80$~cm$^{-1}$ at low $T$ in STO thin films in electric fields \cite{xi3}.
Also, we consider higher-energy polar phonons
like  the TO$_2$- ($170$~cm$^{-1}$)
or the TO$_4$-mode ($545$~cm$^{-1}$).
At the gate/film-interface, such TiO-displacements are
coupled to the holes in the film
\begin{equation}\label{u_s}
H_{{\rm pol}}=\hbar\omega_{TO}\sum_i b_i^{\dag} b_i - \gamma_0 \sum_{i}
  (1-n_i) (b_i^{\dag}+b_i),
\end{equation}
where the phonon operators $b_i^{\dag}$($b_i$) refer to a particular
TiO-mode of energy $\omega_{TO}$,
$\gamma_0=\sqrt{\hbar\omega_{TO}E_p}$ is the hole-phonon
coupling, and $E_p$ is the polaron binding energy.

In the superconducting film, we focus only on the first plane
(ideal confinement) without detailed analysis of the interplanar
charge redistribution. We treat the film in terms of a 2D-Hubbard
model which contains on-site correlations with a nearest
and next-nearest tight-binding dispersion
$\varepsilon_{\ve{k}}=-2t(\cos k_x+\cos k_y)-4t'\cos k_x \cos k_y$:
\begin{eqnarray}\label{hubbard}
H_{{\rm film}}=\sum_{\ve{k}\sigma}\varepsilon_{\ve{k}} c_{\ve{k}\sigma}^{\dag}c_{\ve{k}\sigma}
+U\sum_i n_{i\uparrow} n_{i\downarrow},
\end{eqnarray}
where $c_{\ve{k}\sigma}^{\dag}$ are electron creation operators. We choose the
typical values $t'/t=-0.3$ and $U=8t$. For the doped Cu $3d$ band, the
effective Hamiltonian (\ref{hubbard}) describes the motion of hole singlets
through the lattice of Cu$^{2+}$-ions in CuO$_2$-planes. These spin singlets
are formed by the Cu hole and the O hole of the O-plaquette which is
hybridized with the $d$-states of Cu$^{2+}$ ions.~\cite{zhang}

The combined model (\ref{dg1})--(\ref{hubbard}) now presents a field-driven
two-layer interface system in which the strongly correlated charge carriers in
the film are coupled to TiO-excitons and to phonon states of the gate.
Fig.~\ref{fig1} shows a possible bonding at a CuO$_2$/STO-interface. Here the
oxygen ion is located between Ti$^{4+}$ and Cu$^{2+}$ forming a TiO-Cu bond
with the shortest distance between Cu (film) and O (STO-gate) as compared to
other possible chemical bonding configurations which can appear in
HTSC/STO-heterostructures. The coupling (\ref{dg_s})--(\ref{u_s}) between the
Cu holes and TiO-excitations results in a renormalization of the CuO-hybridization
amplitude $t_{CuO}$ in the planar CuO-squares and, consequently, in a
renormalization of the effective hole hopping parameters
$t=t_{CuO}^2/\varepsilon^O_p$ ($\varepsilon^O_p$ is the energy of the planar O
$p$-state) and $t'$.  In addition, in the configuration shown in Fig.~\ref{fig1}
we also obtain an attractive correction to the one-site repulsion $U$. 
To analyze the resulting electronic parameters in
the film, we first derive an effective Hubbard model by tracing over phonon and
exciton degrees of freedom and subsequently we map this system onto a
$t$--$J$-model with an effective spin exchange energy $J_{{\rm eff}}=4t_{{\rm
eff}}^2/U_{{\rm eff}}$.

\section{Cooperative $p$-$d$-exciton {\rm \&} polaron effect}
We investigate first the TiO hybridization (\ref{dg1})--(\ref{dg_s}). To
eliminate first order coupling terms in $V_{pd}$, we apply the unitary
transformation $U_{\rm exc}=\exp [-\zeta_{pd}\sum_{i\sigma}s_i^y n_{i\sigma}]$
to $H_{pd}+H_{{\rm exc}}+H_{{\rm film}}$ with
$s_i^y=i(d_i^{\dag}p_i-p_i^{\dag}d_i)$,
$\zeta_{pd}=V_{pd}\Delta^0_{pd}/4(\Delta_{pd}^2+2V_{pd}E_{pd})$ and
$\Delta_{pd}=\sqrt{E_{pd}^2+(\Delta^0_{pd}/2)^2}$. Averaging the obtained
expansion over $p$-$d$-exciton states, we find a reduction of the electron
hopping parameters $t$ and $t'$ by the factor $\xi_{pd}=\cos^2 \zeta_{pd}$. For
the considered range of $V_{pd}$, the  attractive corrections $\Delta U_{pd}$
to $U$ are small, $\Delta U_{pd}/U \sim -0.2$, and result in a maximal increase
of $J_{{\rm eff}}/J \sim 1.07$. Consequently, in distinction to weak-coupling
superconductors \cite{kopp}, this contribution of the $p$-$d$ transfer to the
increase of $T_c$ is insignificant. In contrast, a pronounced renormalization
of $J_{{\rm eff}}$ originates from the kinetic term $t_{{\rm eff}}=t\xi_{pd}$.
For $V_{pd}/\Delta^0_{pd} \sim 1$, we find that $t_{{\rm eff}}/t$ can be
reduced by $\xi_{pd}$ down to $\sim 0.7$. On the other hand, the application of
an electric field enhances the exciton factor $\xi_{pd}\rightarrow 1$ and hence
counteracts polaron localization. This may cause a field-tuned delocalization
of holes in the film when we take into account both, $p$-$d$-transfer and
TiO-phonons.

\begin{figure}[b]
\epsfxsize=8.2cm \centerline{\epsffile{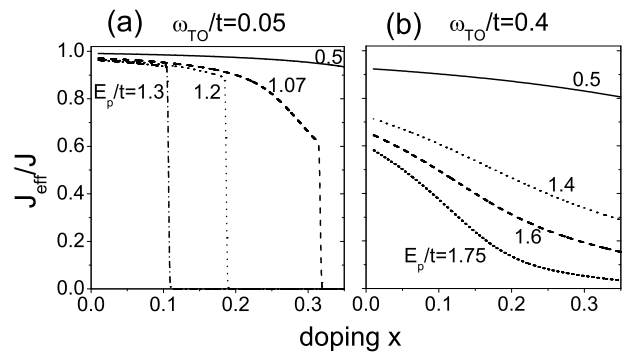}} \caption{Effective exchange
energy $J_{{\rm eff}}$ vs hole doping $x$ for $\Delta_{pd}/4t=3.0$,
$V_{pd}/4t=0.5$, $\varepsilon_g=0$, and different polaron energies $E_{p}$ (in
units of $t$) in the case of coupling to (a) a soft mode
$\hbar\omega_{TO}/t=0.05$; (b) a high-energy mode $\hbar\omega_{TO}/t=0.4$.}
\label{fig2}
\end{figure}

The electron-phonon coupling in $H_{{\rm pol}}$ results in a
polaron effect at the gate/film-interface. With the soft phonon
mode, the typical values for the interface coupling $\gamma_0 \sim
0.01$--$0.1$~eV imply large polaronic energies
$E_p/\hbar\omega_{TO} \sim 0.1$--$5$, where the TiO-mode energy
$\hbar\omega_{TO}/t\sim 10^{-2}$--$10^{-1}$ ($t\approx 0.25$~eV)
can be close to the adiabatic limit. To diagonalize (\ref{u_s}),
we apply a variational Lang-Firsov transformation $U_{{\rm
pol}}(u_i,\alpha,\eta)$ \cite{fehske}. The variational
parameter $\eta$ describes the strength of the polaron effect,
$u_i$ presents interface static distortions, and
$\alpha$ allows for anharmonic excitations. We
assume homogeneity in the low-temperature state of the doped film
and therefore replaced the site-dependent parameters
by their averages: $u_i=u$, $N_{\perp}^{-1}\sum_{i \sigma} \langle
n_{i\sigma}\rangle=1-x$. In the electronic Hamiltonian
$H_{{\rm film}}$, the hopping energies $t$ and $t'$ are reduced by
the polaron band narrowing factor $\xi_{\rm
pol}(\eta)=\exp[-\eta^2\tau {E_p}/{\hbar\omega_{TO}}
\coth\frac{1}{2}\beta\hbar\omega_{TO}]$ where
$\tau=\exp(-4\alpha)$. The parameters $u$, $\eta$, and $\tau$ are
determined by minimization of $\langle U_{{\rm pol}}^{\dag}(H_{{\rm pol}}+H_{{\rm
film}})U_{{\rm pol}}\rangle $, which is averaged over the phonon vacuum
\cite{fehske,roder}. In particular, for $\eta$ we have the
following self-consistent equation:
\begin{eqnarray}\label{gamma}
\eta=1\big/\bigl[1-({\tau}/{\hbar\omega_{TO}})\,e_K(x)\,\xi_{pd}\xi_{\rm pol}\bigr],
\end{eqnarray}
where $e_K(x)\approx -2(t+t')(1-x)$ is obtained from the average kinetic energy
in $H_{{\rm film}}$, which controls the dependence of $\eta$ on $x$. For low $x$,
the contribution of $e_K$ to the denominator in (\ref{gamma}) is significant
which results in small $\eta$ and weak polaron band narrowing $\xi_{\rm
pol}(\eta)$. On the other hand, when the hole-phonon coupling is strong and
$x$ increases, $\eta$ approaches 1, which leads to a localization of holes.
This signifies that the reduced kinetic energy at sufficient concentration of
holes triggers the polaronic localization transition \cite{roder}.
Consequently, the exchange energy $J_{{\rm eff}}$ which is controlled by
$t_{{\rm eff}}=t\xi_{pd}\xi_{\rm pol}$, is drastically suppressed
(Fig.~\ref{fig2}). For the soft TiO-mode
(Fig.~\ref{fig2}(a)), a suppression of $J_{{\rm eff}}$ is obtained for
$E_p/t> 1$ in the relevant range $x<0.3$. Coupling instead to the high-energy
mode (Fig.~\ref{fig2}(b)) increases the minimal values of $E_p$ required for
the localization, and therefore the suppression of $J_{{\rm eff}}$ is substantially
weaker.

\begin{figure}[t]
\epsfxsize=8.4cm \centerline{\epsffile{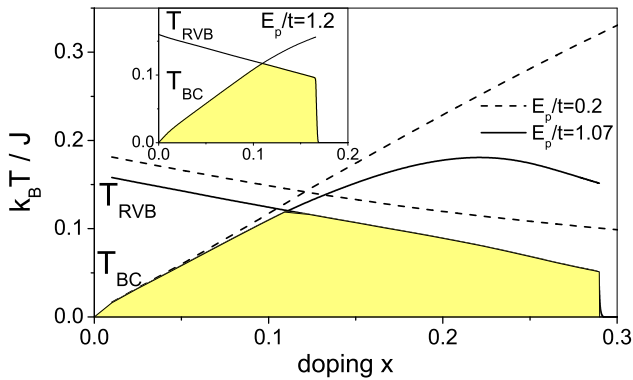}} \caption{
Phase diagram for $d$-wave superconductivity (shaded area) for three
distinct polaron energies. Here
$J/t=0.5$, $V_{pd}/4t=1.0$, $\Delta^0_{pd}/4t=3.0$, and
$\hbar\omega_{TO}/t=0.05$.} \label{fig3}
\end{figure}

\section{Superconductivity in the film}
For our considerations it is
important that for all widely discussed concepts of high-$T_c$
superconductivity, a suppression of $J_{{\rm eff}}$ is feasible due to the loss
of the kinetic energy by interface localization. This suppression may  in fact
result in lower pseudogap temperature $T^* \sim J_{{\rm eff}}$ and critical
temperature $T_c$. In order to relate our findings to a distinct model,  we
approach the superconductivity in the strongly correlated electronic system
within a slave-boson approach \cite{kotliar}. In this scheme spin-carrying
fermions $f_{i\sigma}$ and spinless bosons $h_i$ are introduced  through
$c_{i\sigma}=f_{i\sigma}^{\dag}h_i$ whereby the constraint
$\sum_{\sigma}f_{i\sigma}^{\dag}f_{i\sigma}+h_i^{\dag}h_i=1$ is to be enforced.
We consider the pairing $\Delta_{ij}=(3 J_{{\rm eff}}/4)\sum_{\sigma} \langle
f_{i\sigma} f_{j\bar{\sigma}}- f_{i\bar{\sigma}}f_{j\sigma}\rangle$, bond
$\chi_{ij}=(3 J_{{\rm eff}}/4)\sum_{\sigma} \langle f_{i\sigma}^{\dag}
f_{j\sigma}\rangle$ and boson condensation $\kappa=\langle h_i \rangle ^2$
order parameters. With the mean-field factorization of the $t$--$J$
Hamiltonian, focusing on the uniform solutions $\Delta_{ij}=(-1)^{i_y+j_y}
\Delta_d$, $\chi_{ij}=\chi$, we obtain a free energy, which is to be minimized
with respect to $\Delta_d$, $\chi$, $\kappa$. The system of selfconsistent
equations for the order parameters has been solved numerically. The pairing
amplitude $\Delta_d$ determines the temperature $T_{{\rm RVB}}$, whereas the
boson temperature $T_{{\rm BC}}$ has been estimated according to \cite{kotliar,
wen} from the kinetic energy of the boson pairs assuming a weak
$t_{\perp}/t=10^{-4}$ interplanar coupling in the film. The results are shown
in the form of ($T$, $x$)-phase diagrams in Fig.~{\ref{fig3}} for different
values of the polaron energy $E_p$, where $T_c$ corresponds to the lowest
temperature among $T_{{\rm RVB}}$ and $T_{{\rm BC}}$. In Fig.~{\ref{fig3}}, the
gate field is $\varepsilon_g=0$, and the standard phase diagrams are modified
essentially due to the interface coupling with the soft mode. Here, at low
doping $x<0.12$, the superconducting region is limited by $T_{{\rm BC}}$, and
for larger $x$ by $T_{{\rm RVB}}$. The increase of the polaron energy $E_p$
first affects the overdoped region where the right boundary of the
superconducting phase is shifted from $x_m\approx 0.3$ to $0.17$  as $E_p/t$
increases to $1.2$. With a further increase of $E_p$, the collaps of $T_{{\rm
RVB}}$ limits the superconducting state also in the underdoped region: the
maximally achievable doping, $x_m$, for superconductivity to prevail is
$x_m=0.1$ for $E_p/t=1.3$, already below ``optimal doping''. This is in
striking contrast to the classical HTSC-behavior where $T_c$, after approaching
its maximum at optimal doping, smoothly decreases to $0$ as $x \rightarrow
0.3$. Finally, with $E_p/t>1.4$, the superconducting region disappears due to
the suppression of pairing by the soft-mode excitations in the gate. Note that
for a film containing more than one superconducting plane, the suppression of
pairing is incomplete \cite{multi}.

\begin{figure}[t]
\epsfxsize=8.7cm \centerline{\epsffile{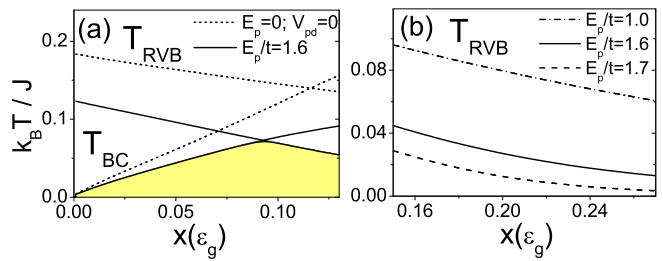}}
\caption{Field-dependent doping $x(\varepsilon_g)$ raised from
initial values (a) $x_0=0$ and (b) $x_0=0.15$: phase diagrams for
$d$-wave superconductivity for different polaronic energies $E_p$;
$V_{pd}/4t=1$, $\Delta^0_{pd}/4t=3$, and $\hbar\omega_{TO}/t=0.4$.} \label{fig4}
\end{figure}

\begin{figure}[t]
\epsfxsize=8.2cm \centerline{\epsffile{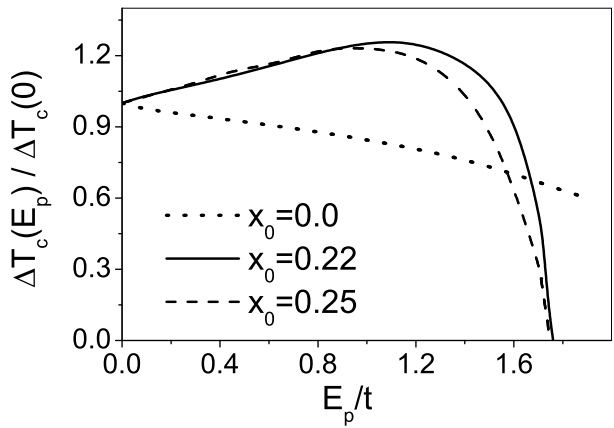}}
\caption{Field-induced shift $\Delta T_c$ vs $E_p$ scaled by $\Delta T_c(E_p=0)$.
Here $V_{pd}/4t=1$, $\Delta^0_{pd}/4t=3$, and $\hbar\omega_{TO}/t=0.4$.} \label{fig5}
\end{figure}

{\it For the field tuned heterostructures},
field-dependent phase diagrams are required. Fig.~{\ref{fig4}}
shows the evolution of such a phase diagram with the increase of
the interface polaron energy $E_p$. The hole doping
$x(\varepsilon_g)$ measures the density of holes injected by
$\varepsilon_g$ into the film which is initially in the insulating
($x_0=0$, Fig.~{\ref{fig4}}(a)) or in the superconducting
($x_0=0.15$, Fig.~{\ref{fig4}}(b)) state. Here,
$x(\varepsilon_g)=Q/N_{\perp}$ with the charge $Q=CV$, the
dielectric capacitance $C=\epsilon_0 \epsilon N_{\perp} a^2/d$,
the lattice constant $a$, the thickness of the dielectric film
$d$, and the bias voltage $V=\varepsilon_g d$. For $x_0=0$, the
maximally achievable  field $\varepsilon_g \approx 10^6$~V/cm for
a gate with $\epsilon=100$ allows to attain a doping of $x \approx
0.13$, which still is in the underdoped range
(Fig.~{\ref{fig4}}(a)). The decrease of $J_{{\rm eff}}$ for larger
$E_p$ leads to the suppression of both $T_{{\rm BC}}$ and $T_{{\rm
RVB}}$. These temperatures also remain strongly dependent on
$\varepsilon_g$. In contrast, for $x_0=0.15$, when doping $x(\varepsilon_g)$
is raised into the overdoped range, the increase of the
polaron energy $E_p$ not merely suppresses $T_{{\rm RVB}}$ (which
limits the superconducting region), but also reduces the slope of
$T_{{\rm RVB}}$ with respect to $x(\varepsilon_g)$.
The latter effect results in a
decrease of the $T_c$ shift, $\Delta T_c=T_{{\rm RVB}}(x_0+\Delta
x(\varepsilon_g))-T_{{\rm RVB}}(x_0)$, where the field
$\varepsilon_g$ injects the hole density $\Delta x=0.03$ into the
film. The corresponding shifts are presented in
Fig.~{\ref{fig5}} and show a strong decrease of $\Delta T_c$
for $E_p/t>1.2$ in the overdoped range, whereas in the underdoped
films $\Delta T_c$ decreases but remains finite. This result
supports the surprising fact that the field does not substantially
change $T_c$ in the overdoped films.

\begin{figure}[t]
\epsfxsize=7.8cm \centerline{\epsffile{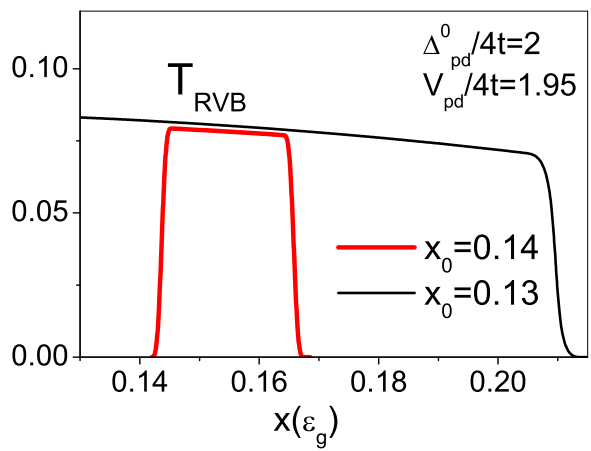}}
\caption{Reentrant behavior of $T_{{\rm RVB}}$ for $E_p/t=1.0$ and
$\hbar\omega_{TO}/t=0.05$.} \label{fig6}
\end{figure}

The hole localization, which is also supported by experiments \cite{eckstein},
presents a serious challenge to current attempts to induce superconductivity in
the strongly correlated films. However, a possible recipe how to delocalize the
holes follows from the analysis of the coupling between the CuO hole and
$p$-$d$-TiO exciton. It was noted before that $\varepsilon_g$ increases the
exciton factor $\xi_{pd}$ and enhances the hole motion for sufficiently large
exciton-hole interaction $V_ {pd}$. To analyze this effect, we investigate the
region $x_0=0.13$--$0.14$ close to the transition into the localized state
where superconductivity is already suppressed, and then apply the field. Due to
the enhancement of $t_{{\rm eff}}$ by $\xi_{pd}(\varepsilon_g)$, for $x_0=0.14$
we find a reentrant transition (Fig.~{\ref{fig6}}) into the superconducting
state. With the injection of more holes by $\varepsilon_g$,  the kinetic term
in (\ref{gamma}) is reduced which leads to a second suppression of
superconductivity at $x\approx 0.17$. For $V_{pd}\approx \Delta^0_{pd}$, the
reentrant superconductivity is stable in a very narrow doping range (see
Fig.~{\ref{fig6}}). For $V_{pd}> \Delta^0_{pd}$, this range rapidly extends
over the full overdoped regime $x\le 0.3$. Such a strong $V_{pd}$, required for
the hole delocalization, can be obtained in designed heterostructures with the
holes in close proximity to the $p$-$d$ TiO-orbitals where Cu-states share an
apical oxygen with the adjacent Ti as shown in Fig.~\ref{fig1}. Alternatively,
we propose to use multilayered cuprates with $3$--$5$ CuO$_2$-planes in a unit
cell to delocalize the holes. Due to the small interplanar distance of $\sim
3.2$~\AA~in some compounds, the polaron-suppressed hopping in the interface
plane is in turn enhanced by the interplanar coupling to the subsequent planes.

We found that the cooperative effect of interface hybridization and lattice
dynamics in strongly correlated superconducting heterostructures
is of crucial importance to their superconducting properties.
The interface-caused suppression of $T_c$ for higher doping is a
possible explanation of the differences in the electric field effect
in overdoped and underdoped SuFETs \cite{mannhart}. The highly
nontrivial behavior of $T_c$ under
a variation of the field allows to propose mechanisms for charge
delocalization which is a subject of further experimental and theoretical studies.

\section*{ACKNOWLEDGEMENTS}
This work was supported through the DFG~SFB-484, BMBF~13N6918A, and DAAD
D/03/36760. We thank J.~Mannhart, G.~Logvenov, P.J.~Hirschfeld and C.~Schneider
for helpful discussions. N.P. thanks U.~Eckern for his continuous support.

\newpage


\newpage


\begin{references}
\bibitem{multi} In an extended description of the film, more CuO$_2$-planes should be coupled by
interplanar electron tunneling. This effectively leads to a
reduction of phonon-induced localization of holes resulting in a
suppressed but finite $T_c$ for larger $x$.
\end{references}
\end{document}